\documentclass{aa}
\input{psfig.sty}
\usepackage{graphicx}
\usepackage{natbib}
\bibpunct{(}{)}{;}{a}{}{,}

\begin{document}
\title{GRB\,050410 and GRB\,050412: are they really dark GRBs?}

\author{ T. Mineo\inst{1}, V. Mangano\inst{1}, S. Covino\inst{2},
G. Cusumano\inst{1},  V. La Parola\inst{1}, E. Troja\inst{1}, P. Roming\inst{3},
D.N. Burrows\inst{3}, 
S. Campana\inst{2},   M. Capalbi\inst{4},   G. Chincarini\inst{2}, 
N. Gehrels\inst{5},   P. Giommi\inst{6},
J.E. Hill\inst{5,7},
F. Marshall\inst{5}, A. Moretti\inst{2}, 
P. O'Brien\inst{8},
M. Page\inst{9}, M. Perri\inst{4},
P. Romano\inst{2},
B. Sbarufatti\inst{1}, 
G. Sato\inst{5}, 
G. Tagliaferri\inst{2} }
\institute{
INAF--Istituto di Astrofisica Spaziale e Fisica Cosmica  di Palermo,
Via Ugo La Malfa 153, I-90146 Palermo, Italy 
\and INAF--Osservatorio Astronomico di Brera, Via E.\ Bianchi 46, I-23807 Merate (LC), Italy
\and Penn State University, 525 Davey Lab, University Park, PA 16802, USA 
\and ASI, Science Data Center, via G.\ Galilei, I-00044 Frascati, Italy % Matteo e Milvia
\and NASA/Goddard Space Flight Center, Greenbelt, MD 20771, USA 
\and ASI, Unit\`a Osservazione dell'Universo, Viale Liegi 26, I-00198 Roma, Italy % Giommi
\and Universities Space Research Association, 10211
Wincopin Circle, Suite 500, Columbia, MD, 21044-3432, USA 
\and  Department of Physics \& Astronomy, University of Leicester, LE1 7RH, UK
\and  Mullard Space Science Laboratory, University College of London, Holmbury
      St Mary, Dorking,  Surrey RH5 6NT
}

\offprints{Teresa Mineo: \\ 
teresa.mineo@ifc.inaf.it}

\date{Received:.; accepted:.}

\titlerunning{}

\authorrunning{T. Mineo et al.}

\abstract{}
{We present a detailed analysis of the prompt and afterglow emission of
GRB~050410 and GRB~050412 detected by Swift for which
no optical counterpart was observed.
}
{We analysed data from the prompt emission detected by the Swift BAT 
and from the early phase of the afterglow obtained by the Swift narrow 
field instrument XRT. 
}
{The $15-150$\,keV energy distribution of the GRB~050410 prompt 
emission shows a peak energy at 53$_{-21}^{+40}$  keV. 
The XRT light curve of this GRB decays as a power law 
with a slope of $\alpha=$1.06$\pm$0.04. The spectrum
is well reproduced by an absorbed power law with a spectral index 
$\Gamma_{\rm x}=2.4\pm0.4$ and a low energy absorption 
$N_{\rm H}$=4$^{+3}_{-2}\times$10$^{21}$ cm$^{-2}$
which is higher than the Galactic value. 
The $15-150$\,keV prompt emission  in GRB~050412 is modelled with a  hard  
($\Gamma$=0.7$\pm$0.2) power law.
The  XRT light curve follows a broken power law with the first slope
$\alpha_1$=0.7$\pm$0.4, the break time  T$_{\rm break}$=254$_{-41}^{+79}$
s and the second slope $\alpha_2$=2.8$_{-0.8}^{+0.5}$. 
The   spectrum is fitted by a  power law with 
spectral index $\Gamma_{\rm x}=1.3\pm0.2$ which is absorbed at low energies by the 
Galactic column. 
}
{The GRB~050410 afterglow light curve  reveals the expected characteristics of
the third component of the canonical Swift light curve. Conversely, a complex 
phenomenology was detected in the GRB~050412 because of the presence of the 
very early break. 
The light curve  in this case  can be  interpreted as being the 
last peak of the prompt emission. 
The two bursts present tight upper limits for the optical emission,
however, neither of them can be clearly classified as dark. For GRB~050410,  
the suppression of the optical afterglow could be attributed
to a low density interstellar medium surrounding the burst. 
For GRB~050412, the evaluation of the darkness is more difficult due to
the ambiguity  in the extrapolation of  the X-ray afterglow light curve.
 }
\keywords{gamma rays:bursts - X-rays: individual(GRB~050410, GRB~050412)}

\maketitle

%%%%%%%%%%%%%%%%%%%%%%%%%%%%%%%%%%%%%%%%%%%%%%%%%%%%%%%%%%%%%%%%%%%%%%%%%%%%%%%%

\section{Introduction}
Gamma-Ray Bursts (GRBs), the brightest explosions in the universe,
produce emission across the whole  electromagnetic spectrum from $\gamma$-rays to radio
wavelengths. However, multi-wavelength observations
 have shown that the optical counterpart  following the prompt emission 
 is detected  in only about 50\% of the well localised events
\citep{depasquale03,roming06,roming+mason06}.

The paucity of optical detections of  GRB afterglows has been explained 
 by invoking different mechanisms
\citep{lazzati02,lamb00,groot98,taylor98,wijers98,totani97}.  More 
recently, by comparing the Swift optical, X-ray and $\gamma$-ray data sets, 
\citet{roming06} identified  a class of  optically ``dark''
GRBs with  higher than normal $\gamma$-ray efficiency. A possible
mechanism proposed for these GRBs is based on a  Poynting flux
dominated outflow \citep{zhang05} where the transfer of the energy  from the
fireball to the medium is delayed, leading to the suppression of the reverse
shock, likely responsible for the prompt optical emission,  and to an apparent high  
$\gamma$-ray efficiency. 
A stronger alternative mechanism proposed is a pure non-relativistic 
hydrodynamical reverse shock 
\citep{kobayashi00, nakar+piran04, beloborodov05, kobayashi05,
 uhm+beloborodov06, macmahon06}.   

The characterisation of optical darkness has been 
previously based  on  
 the upper limit of the optical/NIR  afterglow flux  \citep{rol05,filliatre05,filliatre06}
 or the optical-to-X-ray spectral  index \citep{jakobsson04}. 
In particular, 
if the  optical-to-X-ray spectral index, $\beta_{\rm ox}$, is 
lower than $0.5 $ the afterglow should be classified as dark; while for
\citet{rol05} dark afterglows are those with optical upper limits falling  
below the extrapolation of the X-ray spectrum to the optical range.

The Swift Gamma-Ray Burst Explorer \citep{gehrels04}, successfully launched on
2004 November 20, is a multi-wavelength observatory dedicated to the discovery 
and study of GRBs and their afterglows.  
It carries three instruments: the Burst Alert Telescope 
\citep[BAT;][]{barthelmy05} and the two narrow-field instruments: the X-Ray Telescope 
\citep[XRT;][]{burrows05}  and the Ultra-Violet Optical Telescope 
\citep[UVOT;][]{roming05b}.
Swift's capability of rapidly repointing the spacecraft in a few tens of seconds after a BAT
 detection,  allows a study of the first phases of an afterglow evolution
over a broad energy range from optical to X-rays.

In this paper,  we present the results on the analysis of the prompt
and the afterglow  emission of GRB\,050410 and  GRB\,050412 observed by
Swift, two bursts for which no optical counterpart was detected.

Errors in the paper are relative to a 90\% confidence
level for a single parameter ($\Delta\chi^2$ = 2.71).
Times are referenced from the BAT trigger, T$_0$. 
The decay and spectral indices
are parameterised as follows: $F(\nu,T) \propto T^{-\alpha} \nu^{-\beta}$ where
$F(\nu,T)$ (erg cm$^{-2}$ s$^{-1}$ Hz$^{-1}$) is the monochromatic flux as
a function of time $T$ and frequency $\nu$; we also use $\Gamma=\beta +1$ as
the photon index $N(E) \propto E^{-\Gamma}$ (ph cm$^{-2}$ s$^{-1}$ keV$^{-1}$)
where $N(E)$ is the number of photons at the energy $E$. 

%%%%%%%%%%%%%%%%%%%%%%%%%%%%%%%%%%%%%%%%%%%%%%%%%%%%%%%%%%%%%%%%%%%%%%%%%%%%%%%%

\section{Observation}

GRB\,050410  triggered BAT at 12:14:25.36 UT \citep{fenimore05}.
At the time of the first detection, GRB\,050410 was within the Swift Earth-limb constraint, 
therefore, the observatory executed a delayed automated slew to the BAT
position and started observing
with the narrow-field instruments   $\sim$  32
minutes after the burst.  At that time, the X-ray afterglow was too faint for the XRT to 
produce an on-board centroid, therefore a position  was determined during
ground processing \citep{laparola05}. 
No new source was detected by UVOT within the BAT and XRT error 
circles down to a 3$\sigma$ limiting magnitude of V = 19.9,
B = 21.2 and U =20.9 \citep{boyd05}.
Ground based follow-up also did not detect any optical counterpart for GRB\,050410
down to a limiting magnitude of R $\sim$ 20.5 at times after the burst of
T=13.3 ks \citep{misra05}, T=17.5 ks \citep{ofek05} and T=20.6 ks \citep{rumyantsev05}; 
and limiting magnitudes of Gunn i = 21.5 and Gunn~z = 20.0 at T=56.2 ks  after
the burst \citep{cenko05gcn3231}.
No radio counterpart was  detected  down to  2$\sigma$ upper
limit of 114 $\mu$Jy \citep{soderberg05gcn3223}.

The second burst considered in the paper, GRB\,050412,
was discovered by  Swift at  05:44:03 UT \citep{cummings05,tueller05}. 
The observatory executed an automated slew to the BAT 
position and the XRT and UVOT began taking data 99 s after the BAT
trigger.  The position of the X-ray counterpart was
derived from the first orbit of data and  was given in  \citet{mangano05}.
UVOT data revealed no evidence of a fading source in the 5$\arcsec$  radius XRT error 
circle; the 3$\sigma$  limiting magnitude  was V = 19.1
\citep{roming05gcn3249}. The Chandra X-ray Observatory observed GRB~050412   with the 
ACIS for 20 ks on 2005 April 17, when the source was no longer visible in
the XRT data.  No X-ray source was detected by Chandra and an upper limit on the
unabsorbed $0.5-10$\,keV flux of
3.6$\times$10$^{-15}$ erg cm$^{-2}$ s$^{-1}$ was inferred \citep{berger05gcn3291}.  

Ground based optical follow-up did not detect any counterpart for GRB\,050412
 at early epochs down to a limiting magnitude of 
R $\sim$ 20 at times after the burst of
T=7.1 s \citep{quimby05},
T=250 s \citep{torii05};
T=360 s \citep{cenko05gcn3242} and
T=2.5 ks, \citep{berger05gcn3239}. 
The most constraining optical observation
was performed  with the FOCAS on the Subaru 8.2~m
telescope  atop Mauna Kea
at 2.3 h from the burst and it gave a 3-sigma upper limit of
R$_c$=24.9 on the afterglow emission \citep{kosugi05}.
However, a source at the centre of the XRT
error circle with the magnitude of R$_c$=26.0$\pm$0.5 was detected 7.2 h after
the burst with a second set of observations but because of the 
marginal detection, it was not possible to determine whether  it was a point source
or a galaxy \citep{kosugi05}.
No radio counterpart of GRB~050412 was detected down to 2$\sigma$ upper limit
of 38$\mu$Jy \citep{soderberg05gcn3277}.

%%%%%%%%%%%%%%%%%%%%%%%%%%%%%%%%%%%%%%%%%%%%%%%%%%%%%%%%%%%%%%%%%%%%%%%%%%%%%%%%

\section{BAT and XRT data reduction and analysis}

The BAT event data  were re-analysed using the standard analysis 
software\footnote{see http://swift.gsfc.nasa.gov/docs/swift/}
included in the HEAsoft 6.0.4 package.
Response matrices  were generated with the task {\small {\sc BATDRMGEN}} using the
latest spectral redistribution matrices. 
The BAT background was subtracted using a mask-weighting technique that is only
effective up to 150 keV.

XRT data  were  calibrated, filtered
and screened with the XRTDAS  software package, included in the HEAsoft 6.0.4
package, to produce cleaned photon
list files. Only time intervals with the CCD temperature below $-$47~C$^o$
were selected.    
Table~\ref{tab1}  shows the observation
log for  GRB~050510 and GRB~050512.

\begin{table}[ht]
\caption{Observation log for the data used in the analysis of
  GRB~050510 and GRB~050512.  }
\label{tab1} 
\begin{center}
\begin{tabular}{cccr}
\hline
Obs. \# & \multicolumn{1}{c}{Start Time (UT)} & \multicolumn{2}{c}{Exposure} \\
                  & (yyyy-mm-dd hh:mm:ss) &   WT (s) & \multicolumn{1}{c}{PC (s)} \\
\hline
\multicolumn{4}{l}{{\it GRB~050410 }} \\
1  & 2005-04-10~12:46:24 & 9\,436 & 8\,635 \\
2  & 2005-04-12~16:15:47 &  --  & 4\,740\\
3  & 2005-04-12~00:11:47 &  --  & 1\,473\\
4  & 2005-04-14~00:22:44 & --   & 3\,184\\
5  & 2005-04-15~00:30:44 & --   & 10\,895\\
6  & 2005-04-16~00:32:01 & --   & 13\,086\\
7  & 2005-04-17~16:45:01 & --   & 6\,418\\
8  & 2005-04-19~00:50:42 & --   & 9\,832\\
9  & 2005-04-20~00:54:29 & --   & 11\,123\\
10 & 2005-04-21~00:55:58 & --   & 1\,828\\
%\hline
 & Total exposure (s)             & 9\,436 &  71\,214   \\
&  &  & \\
\multicolumn{4}{l}{{\it GRB~050412 }} \\
1  &2005-04-12~05:45:47  & 3\,019 & 1\,950 \\
2  &2005-04-14~00:04:59  & --   & 3\,719 \\
3  &2005-04-17~09:28:35  & --   & 6\,200 \\
 & Total exposure (s)       & 3\,019 &11\,869 \\   
\hline
\end{tabular}
\end{center}
\end{table}

Standard grade selection, 0--12 for Photon Counting (PC)
mode  and 0--2 for Window Timing (WT) mode \citep{hill04},  was used
for both spectral and timing analysis.  
Ancillary response files for PC
and WT spectra were generated with the standard {\small {\sc XRTMKARF}} 
tool (v 0.5.1) using calibration files from CALDBv.2.3. 

WT data were extracted in a rectangular region 40$\times$20 pixels along the image
strip which included about 96\%  of the Point Spread Function (PSF).
PC data were  extracted from a circular region of 20 pixels radius for 
observing intervals with a  rate lower than 0.2 counts s$^{-1}$.   
In time periods with higher rates, the extraction region used was 
 an annulus with  an inner radius of 3 pixels and
an outer radius of 20 pixels in order to take into account  pile-up effects. 
The two regions included, assuming an  average energy of 1.5 keV,
 92\% (circular) and 48\% (annulus) of the PSF, respectively. 
The WT and PC backgrounds   were extracted from regions far from the GRB counterpart
and from  any sources in the field.
WT and PC spectral analysis of  both GRBs was performed in the energy range $0.7-10$\,keV
to avoid  residual background due to bright Earth contamination and dark
current that are more dominant at lower energies.

\subsection{GRB~050410}
GRB\,050410  triggered  Swift twice  as a result of an anomalous
time-out of the triggering code  \citep{fenimore05}.
In our analysis, we consider as reference  time the 
 time of the first trigger (T$_0$=12:14:25.36 UT).  

\begin{figure}
\vbox{
\psfig{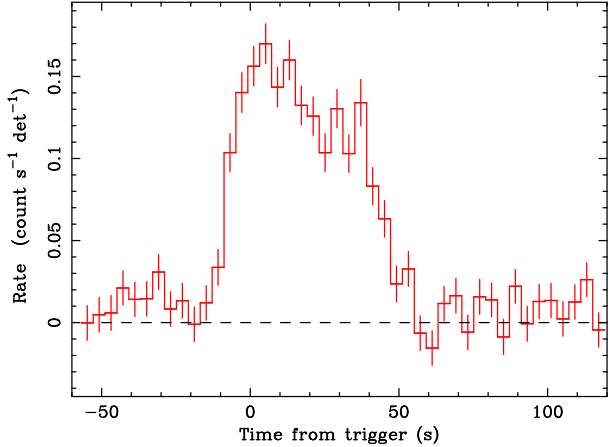}
}
\caption{GRB~050410: BAT light curve in the energy range $15-150$\,keV. Times are
 referenced to the first burst trigger (12:14:25.36 UT). }
\label{grb050410_lc_bat}
\end{figure}

The $15-150$\,keV light curve of the  prompt emission,
shown in  Fig.~\ref{grb050410_lc_bat},  is characterised by a broad
peak with  T$_{90}$=44$\pm$1 s. 
The burst shape can be modelled by two Gaussians of the same width 
(11$\pm$2 s), the first centered at 5$\pm$1 s and the second at 32$\pm$2 s.  

The  $15-150$\,keV energy distribution of the  burst, 
modelled  with a simple power law 
of photon index $\Gamma=1.66\pm0.07$,  
gave a marginally acceptable fit with a
$\chi^2_{\rm red}$ of 1.67 for 36 dof (see Table~\ref{tab3}). 
Better fits  were obtained using a cut-off
power law or a Band model \citep{band93}  with the high energy
photon index, $\beta \, _{\rm Band}$, 
fixed to $-$2.3 according to the expected value \citep{preece00}. The best  fit values 
of the spectral parameters are shown in Table~\ref{tab3}.
The improvement with respect to the power law fit is significant
for both models, with a  chance random probability of 
$\sim$7$\times$10$^{-5}$, assuming that the errors are normally distributed.
The fluence in the $15-150$\,keV energy range was (4.6$\pm$0.1)$\times$10$^{-6}$ erg cm$^{-2}$.

\begin{table}[htb]
\caption{Best fit parameters for the spectral analysis  in the energy ranges
   $15-150$\,keV (BAT) and $0.7-10$\,keV (XRT) for GRB~050410 and GRB~050412. }
\label{tab3}
\begin{center}
\begin{tabular}{lcc}
\hline
\hline
                      & GRB 050410     & GRB 050412  \\
\hline
\multicolumn{3}{c}{$15-150$\,keV (BAT)} \\
 \multicolumn{3}{l}{\it Power Law}   \\          
 $\Gamma$                  & 1.66$\pm$0.07 & 0.7$\pm$0.2  \\
 $\chi^2_{\rm red}$ (dof)      & 1.67  (36)    & 0.55 (15) \\
                           &                    \\
\multicolumn{3}{l}{\it  Band Model with $\beta \, _{\rm Band}$=$-$2.3 }\\
$\alpha_{\rm Band}$        & $-$0.79$\pm$0.09   &  --  \\
 $E_p$  (keV)              & 53$_{-21}^{+40}$  &  --  \\
 $\chi^2_{\rm red}$ (dof)      & 1.11  (35)  &  -- \\
                           &                    \\
 \multicolumn{3}{l}{\it  Cut-off Power Law }\\
 $\alpha_{cutoff}$         & $-$0.8$\pm$0.3   &  -- \\
 $E_p$                     & 54$_{-17}^{+31}$  &  --  \\
 $\chi^2_{\rm red}$ (dof)      & 1.08  (35)  &  \\
                           &                    \\
\multicolumn{3}{c}{ $0.7-10$\,keV (XRT)} \\
\multicolumn{3}{l}{\it  Power Law}    \\  
 $N_{\rm H}$ (cm$^{-2}$)  &4$^{+3}_{-2}\times$10$^{21}$ & 2.21$\times$10$^{20}$  \\     
 $\Gamma_{\rm x}$              & 2.4$\pm$0.4  & 1.3$\pm$0.2  \\
 $\chi^2_{\rm red}$ (dof)  & 0.64  (13)  &  1.18  (17) \\
\hline
\end{tabular}
\end{center}
\end{table}

The XRT observed  GRB\,050410  for 10 orbits for a total
exposure  of $\sim$9 ks in WT mode and  $\sim$71 ks in PC mode. 
The position of the burst, determined with astrometry solutions,
was RA$_{\rm J2000}=$05$^{\rm h}$  59$^{\rm m}$  13$^{\rm s}$.94 \,
Dec$_{\rm  J2000}=+$79\degr \, 36\arcmin \, 11\farcs7, with an 
uncertainty of 2\farcs3 \citep{butler07}. 
These coordinates lie outside the error circle (3\farcs7 radius)
of the position derived with the tool {\small {\sc XRTCENTROID}}
and  42$\arcsec$ from the BAT  position reported by
\citet{fenimore05}.

\begin{figure} 
\centerline{
\psfig{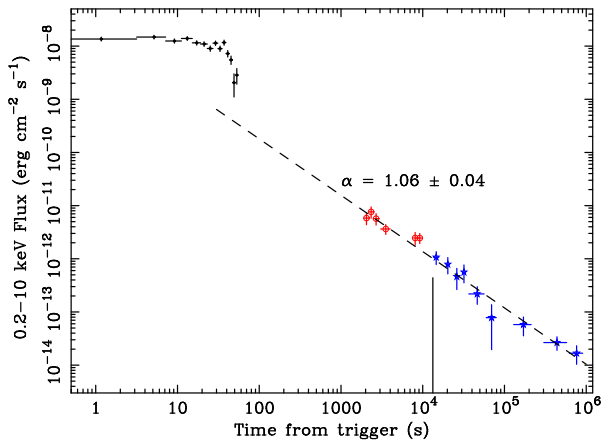}}
\caption{GRB050410 light curve. The XRT $0.2-10$\,keV  count rate was
  converted to flux by applying a conversion factor derived from the spectral
  analysis. The dashed line  represent the best fit model. The BAT light curve
  was extrapolated to the XRT band by converting the BAT count rate with a factor
  derived from the Band spectral parameters (see Table 2). The vertical line
  at 13.3 ks indicate the time of the \citet{misra05} optical upper limit.
}
\label{lc_bat+xrt}
\end{figure}

The GRB~050410 XRT light curve shows a clear decay  with time that 
can be well modelled ($\chi^{2}_{\rm red}$=0.87
with 13 d.o.f.) by a single power law with a slope $\alpha=$1.06$\pm$0.04.
The  light curve, shown in  Fig.~\ref{lc_bat+xrt}, was converted 
to  flux in the  $0.2-10$\,keV band using a
conversion factor derived from the best fit spectral model (see below). 
In the same figure, 
the BAT light curve extrapolated to the  $0.2-10$\,keV  energy range is also plotted.  
The extrapolation was obtained by multiplying the $15-150$\,keV
rate by the average rate-to-flux conversion factor for GRB~050410
derived from  the Band model parameters. 

WT and PC spectra were fitted simultaneously because no  spectral
evolution with time was detected. The two spectra normalisations 
were left free to vary, thus accounting for the different average flux levels 
in the  two  observing periods. An absorbed power law  reproduces well the emission
($\chi^2_{\rm red}$ of 0.64 for 13 d.o.f.) with the best fit parameters (see
also Table~\ref{tab3})
$\Gamma_{\rm x}=2.4\pm0.4$ and $N_{\rm H}$=4$^{+3}_{-2}\times$10$^{21}$
cm$^{-2}$.
Note that the absorbing column is non-zero at the 4-sigma level, 
and the best-fit value of the absorbing column is $\sim$ 5 times larger than 
the Galactic one \citep[7.53$\times$10$^{20}$ cm$^{-2}$;][]{dickey90}.

%%%%%%%%%%%%%%%%%%%%%%%%%%%%%%%%%%%%%%%%%%%%%%%%%%%%%%%%%%%%%%%%%%%%%%%%%%%%%%%%

\subsection{GRB~050412 }
The GRB~050412   prompt emission showed a double-peaked
 structure with the first peak  centered at  $-$1.1$\pm$0.4~s from the trigger  
(width  of 4.7$\pm$0.6 s),  and the second  fainter and wider (11$\pm$3 s)
 peak centered at 16$\pm$5 s (see Fig.3). The estimated T$_{90}$ in the $15-150$\,keV 
energy band was 27$\pm$1 s. 
The hardness ratio  ($50-150$\,keV and $15-50$\,keV)
does not show any significant evidence of spectral variation  during the
 burst evolution.

The $15-150$\,keV energy distribution was well described  ($\chi^2_{\rm red}=0.55$ for 15
d.o.f.) by a single power law with an hard photon index of
$\Gamma=0.7\pm0.2$.  The T$_{90}$ burst fluence in the $15-150$\,keV energy range 
was (6.3$\pm$0.3)$\times$ 10$^{-7}$ erg cm$^{-2}$.
A Band model and a cut-off power law were not able to reproduce
 this spectrum because some of the fitting parameters are
 not constrained.

\begin{figure}
\psfig{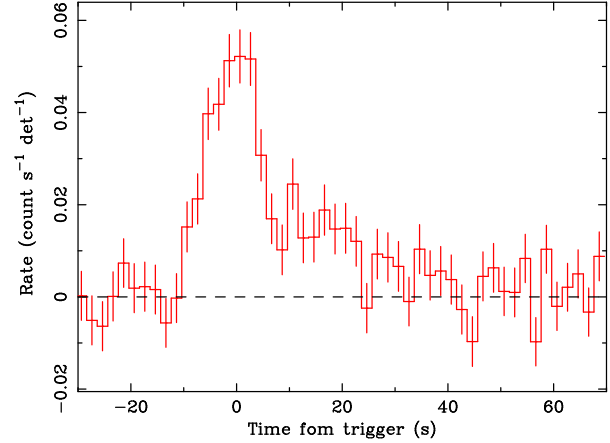}
\caption{GRB050412: BAT light curve in the energy range $15-150$\,keV.  }
\label{lc_bat}
\end{figure}

The XRT detected a  rapidly fading  source at 
RA$_{\rm J2000}=$12$^{\rm h}$  04$^{\rm m}$  25$^{\rm s}$.2 \,
Dec$_{\rm  J2000}=\,$-01\degr \, 12\arcmin \, 00\farcs4,
with an uncertainty of 4\farcs2.   
This position, derived with the tool {\small {\sc XRTCENTROID}},
 is 60\farcs8 from the BAT position \citep{tueller05}  and it is in agreement 
with the refined value derived with astrometry technique \citep{butler07}. 

\begin{figure}
\centerline{
\psfig{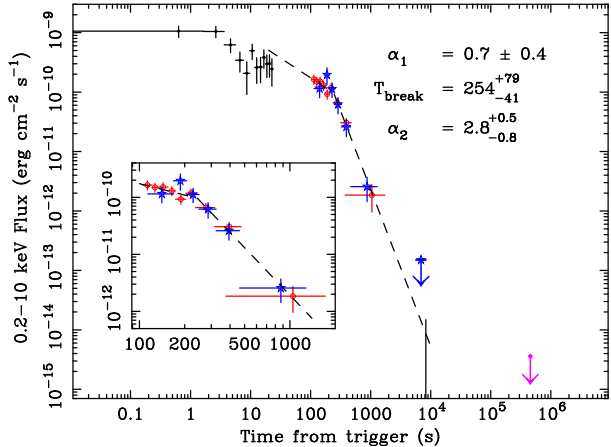}}
\caption{GRB050412 light curve. The XRT $0.2-10$\,keV count rate was
converted to flux by applying a conversion factor derived from the spectral
analysis. The red circles  indicate WT data, the blue stars indicate PC
data; the solid line represent the best fit model. The BAT light curve
was extrapolated to the XRT band by converting the BAT count rates with the factor
derived from the spectral parameters. The 3$\sigma$ upper limit at 4.5$\times$10$^{5}$ s 
was obtained  by Chandra. The vertical line
  at 8.3 ks indicate the time of the \citet{kosugi05} optical upper limit.
}
\label{lc_bat+xrt_12}
\end{figure}

The XRT light curve  is not consistent with a single power law
($\chi^2_{\rm red}=2.3$ with 18 d.o.f.). 
A better fit ($\chi^2_{\rm red}=0.7$ with 14 d.o.f.) can be obtained with a 
broken power law whose best fit parameters are:
 $\alpha_1$=0.7$\pm$0.4, 
$\alpha_2$=2.8$_{-0.8}^{+0.5}$ and T$_{\rm break}$=254$_{-41}^{+79}$ s.
The GRB~050412 light curve was converted 
to the $0.2-10$\,keV  flux with a procedure similar to the one used for GRB~050410, and
plotted in Fig.~\ref{lc_bat+xrt_12} together with
the BAT light curve  extrapolated to the XRT energy range. 
In the same figure, 
the 3$\sigma$ upper limit obtained  by  Chandra is also plotted \citep{berger05gcn3291}. The upper
limits derived from Swift observations 2 (2005-04-14) and 3 (2005-04-17) are not
plotted in the figure, since they are less constraining than the Chandra one.

 The first orbit of the GRB~050412 observation is affected by a continuous
switching between PC and WT mode (caused by flickering pixels produced by the
 high CCD temperature and by the bright Earth contamination); 
thus the two operational mode spectra are  almost contemporaneous. 
The WT spectrum only employed the first 1400 s of data
to avoid  a background flare present in the second part of the observation.
WT and PC spectra were fitted simultaneously with an absorbed power-law model,
 leaving the normalisation free in order  to take into account differences in the flux levels.
The measured absorption column was  consistent  with
the Galactic  \citep[2.21$\times$10$^{20}$ cm$^{-2}$;][]{dickey90} and was  
fixed to this value.
The best fit power-law spectral index was 
$\Gamma_{\rm x}=1.3\pm0.2$  with a reduced $\chi^2$ of 1.18 (17 d.o.f.). 
Results of the spectral analysis are shown in  Table~\ref{tab3}.

%%%%%%%%%%%%%%%%%%%%%%%%%%%%%%%%%%%%%%%%%%%%%%%%%%%%%%%%%%%%%%%%%%%%%%%%%%%%%%%%
\section{Discussion}

The discussion on the results of GRB~050410 and GRB~050412 analysis 
is based on two main points: (i) the comparison of the X-ray afterglow light curves
with other Swift results,  particularly important for GRB~050412 where
a very early break ($\sim$250 s) followed by a steep decay 
$\alpha_2$=2.8$_{-0.8}^{+0.5}$ were detected; 
(ii) the classification  of  the two bursts as dark.

\subsection{The X-ray light curve}
The X-ray afterglow light curves of many bursts
manifest a similar behaviour \citep{obrien06,nousek06,willingale06}.
The canonical X-ray light curve has an initial  steep decay
(usually interpreted as emission from the tail of the
prompt GRB),  followed by a flatter decay phase that can last up to 10$^5$ s 
(interpreted as a refreshing of the forward shock), and  a  final
steeper decay phase with power law indices consistent with the values measured 
before the launch of Swift \citep{frontera03}. 
This  canonical  light curve 
is  consistent with about  60\% of the Swift afterglows.  In many of the
non-conforming cases, 
the first, or the first and the second branches are missing. The lack of 
detection of these portions of the light curve
cannot  always be attributed to  missing  observational data.
Moreover, in about half of the afterglows, late X-ray
flares,  probably  due to continued activity of the central
engine,  are observed \citep{obrien06, nousek06, zhang06}. 

The light curve of GRB 050410, starting from about 2000 s from the trigger,
shows a constant decay slope  that can be interpreted as the third branch of the
canonical light curve. The synchrotron radiation theory  applied to the 
fireball model  predicts that the temporal decay index,
$\alpha$, of a GRB afterglow and the spectral slope, $\beta$,   are linked by relations that
depend (i) on the density profile of
the external medium  \citep{meszaros97, chevalier99}; 
(ii) on the observer's perception of the geometry of the 
expansion: spherical whenever the expansion velocity corresponds to a 
Lorentz factor, $\Gamma_{\rm e}$, such that $\theta_0>\Gamma_{\rm e}^{-1}$,
and beamed for $\theta_0<\Gamma_{\rm e}^{-1}$
with $\theta_0$ being the half opening angle of the jet  \citep{sari99,rhoads99}; 
(iii) on the observational frequency and its relation to the typical
synchrotron frequency of newly shocked electrons, $\nu_m$, and to the cooling
frequency $\nu_c$ corresponding to a synchrotron cooling time equal to the
hydrodynamical expansion time  \citep{sari98}.
In particular, for fireball expansion in a uniform interstellar medium we are
usually in the $\nu_x>\nu_c$ regime and expect $\alpha=(3p-2)/4$ and 
$\beta=p/2$, with the energy distribution of the electrons, $p$, greater than
2, before the jet edge effect is manifested. 
This relation is satisfied by  GRB 050410 with a value  $p=2.08\pm0.05$. 
This value is lower than $p$ values predicted by numerical simulations 
(2.2-2.5),  but still consistent for  particle acceleration at ultraluminous
shocks.
 From this result, we can derive a lower limit of 10 d to the time after
  the trigger of any jet break.
The XRT light curve of GRB 050410 is therefore consistent with a 
pre-Swift X-ray afterglow light curve, or with the third branch of the
canonical Swift afterglow light curve. The late start of XRT observations 
($\sim$2000 s after the trigger) might have prevented the detection of the
previous branches.

According to the  typical Swift afterglow behaviour, the early break in the X-ray
light curve of GRB~050412  may be interpreted as the flat to steep
transition corresponding to the end of the refreshing of the forward shock.
However the measured spectral index $\beta_{\rm x}$=0.3$\pm$0.2 would imply
a flat electron energy distribution with  $p=0.7\pm0.3$.
The expected temporal slope, either for spherical expansion in a uniform
medium or for a standard isotropic wind model in the $p<2$ and $\nu_x>\nu_c$
regime, is inconsistent with $\alpha_2$. 

The values of $\alpha_1$ ($0.7\pm0.4$),  $\alpha_2$ ($2.8_{-0.8}^{+0.5}$) and 
$\beta_{\rm x}$ ($0.3\pm0.2$) of GRB~050412 fit the closure relations corresponding to
a jet break occurring
 in the $\nu_m<\nu_x<\nu_c$ regime for $p=(2 \beta_{\rm x}+1)=1.7\pm0.3$ with 
$\alpha_1=3 \, (2 \beta_{\rm x}+3)/16$ and $\alpha_2=(2 \beta_{\rm x}+7)/4$
according to \citet{dai01}. An early  achromatic  break has also
been detected  in the X-rays and optical light curve  for GRB 050801  \citep{rykoff06,covino06},
while  achromatic breaks  usually occur at T$>$10$^{4}$ s
 \citep{blustin06,romano06, soderberg06, bloom03, frail01,covino06}. 
However, in the case of GRB 050412, this explanation is very unlikely
  because such an early time for the temporal break
would require a very small value for the jet angle ($\theta_0 < 1$ deg)
which is at variance with the beaming angle ($\theta_0 > 3$ deg) 
 computed,  assuming  a peak energy greater than 150 keV, from the
\citet{amati06} and \citet{ghirlanda04} relations, 
that holds for most of the bursts.

Alternatively, the first part of the XRT light curve 
can  be considered  as the  last peak of the prompt emission, 
and the decay rate beyond 250 s
can be interpreted as due to curvature effect after an
instantaneous turn off of the source  \citep{kumar00}. Within this model the  relation that
should be satisfied  is $\alpha_2=\beta_{\rm x}+2$ which is in agreement with our results.
 After the final  detection at 1 ks, the X-ray light curve may start following
a typical afterglow decay without violating the later upper limits,
however a drop of more than three order of magnitudes in flux within the first 
10$^4$ s from the trigger and before the emergence of the afterglow is
uncommon to Swift GRBs. It is then possible that  
GRB~050412 was  a burst without afterglow (a naked burst) as GRB~050421 \citep{godet06}.

\subsection{Darkness}

GRB\,050410 and GRB\,050412 do not have identified optical/infrared  
afterglows,  therefore, it is worth considering how dark these bursts  are. 
Both bursts are located in regions with moderate or small  Galactic extinction 
($E_{B-V} \sim 0.11$ and $0.02$ for GRB\,050410  and GRB\,050412,
respectively). 
The available optical upper limits for GRB\,050410 are too shallow for any
meaningful claim,  as shown in Fig. 5 where the broad band 
spectral energy distribution
at the time of the \citet{misra05} optical measurement is plotted. 
The  comparison with the X-ray flux ($\beta_{\rm ox} 
\sim 0.8$ computed at 11 h from the burst) indicates that it  was a rather 
regular burst regards to its gamma-ray to X-ray ratio 
 \citep[see also][]{roming06}. 
However, the 2-10 keV flux computed at 1 h (2.4$\times$ 10$^{-12}$ erg cm$^{-2}$
s$^{-1}$) and at 11 h (1.9$\times$ 10$^{-13}$ erg cm$^{-2}$ s$^{-1}$)
shows that this burst has a relatively faint X-ray afterglow 
\citep[see Fig.4 in][]{roming06}. 

\begin{figure}
\centerline{
\psfig{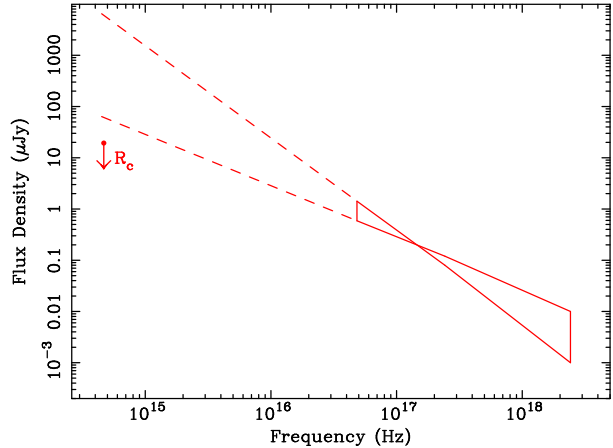}}
\caption{Spectral energy distribution of the GRB050410 at 13.3 ks from the
burst. The optical measurement is the upper limit reported in
\citet{misra05}; the X-ray spectrum is computed evaluating the 0.2-10 keV flux
at the time of the optical point 
and assuming as spectral model the power law reported in Table 2.}
\label{sed050410}
\end{figure}

A possible interpretation of a low afterglow 
flux level is a low density medium \citep{groot98,frail99,taylor00}. 
 Assuming that the low X-ray flux is due to a low-density circum-burst medium, we 
derived  an estimate of the
interstellar medium near the GRB under some reasonable assumptions.
From the detected peak energy, $E_p$, in the BAT spectrum, and using the Amati relation
\citep{amati06} the values of the  energy of the afterglow $E_a$ ($\sim$
2$\times$10$^{51}$ erg) and of  the redshift $z$ ($\sim$ 0.4) were inferred. 
Assuming an electron index of $p=2.5$, 
taking into account the decay of the afterglow and assuming 
that the observing frequency is between the peak
frequency $\nu_{\rm m}$ and the cooling frequency $\nu_c$, we derived  $n<0.01$
cm$^{-3}$.
The estimate of $n$, subject to the uncertainty on $z$, $E_a$ and $p$ (which increases
with $z$ and decreases with $E_a$ and $p$) is lower than the typical value
 of  GRBs with  optical afterglow  ($\sim$ 1 cm$^{-3}$),
but similar to values which have  already been derived for other GRBs
\citep{panaitescu02}.

\begin{figure}
\centerline{
\psfig{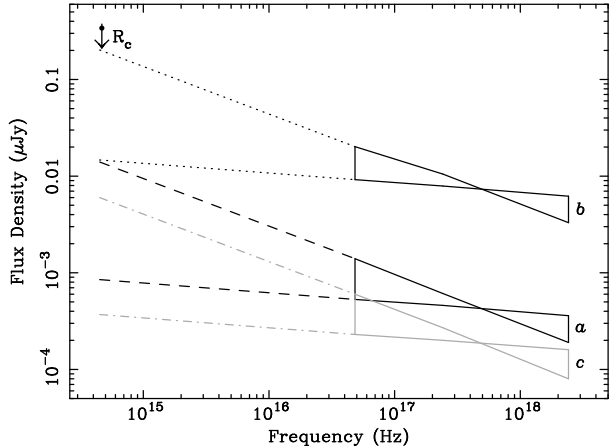}}
\caption{Spectral energy distribution of the GRB050412 at 2.3 hr from the
burst. The optical measurement is the upper limit reported in
 \citet{kosugi05}
The X-ray spectrum labelled with {\it a} is computed extrapolating the XRT 
with a power law with temporal index 2.8 and assuming no spectral variation during the 
burst evolution (see Table 2). The spectrum labelled with {\it b}  assumes that the X-ray
 light curve become flat with a flux value equal to the upper limit 
measured by XRT at 7 ks from the burst; the curve  labelled with {\it c} is computed
assuming that the light curve become flat at the level of the Chandra
upper limit \citep {berger05gcn3291}.
}
\label{sed050410}
\end{figure}

 Things are considerably different in the case of GRB 050412:
about 2.3 hr after the burst the optical afterglow was not  detected with 
an upper limit of $R_c = 24.9$ (Kosugi et al. 2005). In the \citet{rol05} Fig.\,2 the
GRB\,050412 representative point would immediately  classify it as  a very
  dark burst. 
However, while any possible optical afterglow for this event is undoubtedly 
faint already at about 2 hr after the high energy event, to discuss the darkness 
of the burst a  comparison with the detected X-ray flux is mandatory. 
The X-ray light curve of  GRB\,050412 shows a break at about  4
minutes  after the burst and then a steep decay with a temporal index of about
2.8  (see Sect.\,3.2). In computing the optical-to-X-ray spectral index
at 11 hr after the burst,we may either assume  that the optical upper limit  went on 
fading with $T^{-1}$ \citep[as assumed in][]{jakobsson04}, or that 
the optical decay follows the X-ray light curve.  
In a similar way, the X-ray density flux computed  at 11 hr from the
burst can be evaluated from the extrapolation of the decay curve, or assuming
that, at that time, the burst flux reached the level of the Chandra upper limit.
The  resulting optical-to-X-ray spectral index 
 $\beta_{\rm ox}$   would not clearly identify this  event
as dark because it ranges from $\sim$0.5 to $\sim$1 depending on how
the optical and X-ray flux extrapolation are performed.
Of course, the brighter  the extrapolated 
X-ray flux, the more  convincing is the classification of GRB\,050412 as a 
dark  event.

 Figure 6 presents the optical upper limit by \citet{kosugi05} at 2.3 hr 
from the trigger together with the  X-ray spectrum  computed extrapolating the XRT 
flux with the measured decay rate ($\alpha_2$=2.8) and assuming no spectral
variation  with time (see label  {\it a}).  
In the same figure, the X-ray spectrum is also computed assuming that 
the flux at 2.3 hr was as high as the upper limit measured by XRT at 7 ks from the
burst (see label {\it b}) and that
the light curve became flat at the level of the Chandra
upper limit \citep[see label {\it c};][] {berger05gcn3291}.
However, we have seen in section 4.1 that the 
most likely interpretation of the XRT light curve of GRB\,050412 is not an
afterglow, but the tail of the prompt emission from curvature effect. 
If this is the case, the possible underlying
X-ray afterglow at 2.3 hr is expected to be 
softer than the measured
spectrum (with a photon index $\Gamma \sim$2).
 This would make the  X-ray spectra in Fig. 6 steeper, bringing
the expected  optical emission well above the observed upper limit,  but it would
not lead to a firm conclusion about the darkness of GRB~050412.

As last point, looking at Fig.\,3 in
\citet[][]{roming06}  both GRB\,050410 and GRB\,050412 appear fairly normal 
considering the  X-ray flux at 1 hour compared to their prompt  $\gamma$-ray flux.

\section{Conclusion}
We have presented a detailed analysis of the prompt and afterglow emission of
 GRB~050410 and GRB~050412 detected by Swift.  
For either burst no optical counterparts were detected.
Results of the analysis can be summarised as follow:
\begin{itemize}
\item
The  prompt emission lasted  44$\pm$1 s with a $15-150$\,keV  fluence 
of 4.6$\pm$0.1$\times$10$^{-6}$ erg cm$^{-2}$ for the first burst (GRB~050410) 
and  27$\pm$1 s  with a  $15-150$\,keV fluence of
(6.3$\pm$0.3)$\times$ 10$^{-7}$ erg cm$^{-2}$ for the second burst (GRB~050412).
The $15-150$\,keV average energy distribution
of the GRB~050410  emission was fitted by a Band model with the
peak energy at 53$_{-21}^{+40}$  keV and a low energy slope of
$-$0.79$\pm$0.09 after fixing the high energy slope  to $-$3.  
The GRB~050412 $15-150$\,keV emission  was modelled with a  hard  
($\Gamma$=0.7$\pm$0.2) power law suggesting a peak energy  above  BAT energy range.
\item
The GRB~050410 XRT light curve can be modelled with a single  power law 
 with a slope of $\alpha=$1.06$\pm$0.04. The average spectrum
is  reproduced by an absorbed power law with a spectral index 
$\Gamma_{\rm x}=2.4\pm0.4$ and  low energy absorption 
$N_{\rm H}$=4$^{+3}_{-2}\times$10$^{21}$ cm$^{-2}$ which is
higher than the galactic value. 
\\
The GRB~050412  XRT light curve  follows a broken power law with the first slope
 $\alpha_1$=0.7$\pm$0.4, the break time  T$_{\rm break}$=254$_{-41}^{+79}$
 s and the second slope $\alpha_2$=2.8$_{-0.8}^{+0.5}$. 
The average spectrum was fitted by a  power law with a
spectral index $\Gamma_{\rm x}=1.3\pm0.2$  and absorbed at low
energies by a column consistent  with
the Galactic  ($N_{\rm H}$=2.21$\times$10$^{20}$ cm$^{-2}$). 
\item
The GRB 050410 afterglow light curve manifests the expected characteristics of
the third component of the canonical Swift light curve
and can be interpreted as that X-ray afterglow of a spherical fireball
expanding in a uniform medium. In contrast, a rather  complex 
phenomenology was detected in the GRB~050412  X-ray light curve
 because of a  very early break ($\sim$250 s). 
A possible explanation for the observed phenomenology
suggests the detection of a tail of the prompt emission.  
\item
Upper limits exist for the afterglows of both bursts in the optical, and
the upper limit is quite severe in the case of GRB 050412. However, neither
burst can be clearly classified as a dark burst according to the
definition given by \citet{jakobsson04}.  GRB~050410 has a $\beta_{ox}=0.8$ and
the suppression of the optical afterglow could be attributed
to  a low density of the interstellar medium surrounding the burst. 
For the second burst, the  proper evaluation of the $\beta_{ox}$ is quite difficult due to
the ambiguity  in the extrapolation of  the X-ray  light curve.
\end{itemize}

%%%%%%%%%%%%%%%%%%%%%%%%%%%%%%%%%%%%%%%%%%%%%%%%%%%%%%%%%%%%%%%%%%%%%%%%%%%%%%%%

\begin{acknowledgements}
This work is supported at INAF by ASI grant I/R/039/04
and by MIUR grant 2005025417,
at Penn State by NASA contract NAS5-00136 and
at the University of Leicester by PPARC.
We gratefully acknowledge the contributions of dozens of members of the XRT and UVOT team at
OAB, PSU, UL, GSFC, ASDC, and MSSL and our subcontractors, who helped make this mission possible.
\end{acknowledgements}

%%%%%%%%%%%%%%%%%%%%%%%%%%%%%%%%%%%%%%%%%%%%%%%%%%%%%%%%%%%%%%%%%%%%%%%%%%%%%%%%

\bibliographystyle{aa}
\bibliography{mt6594}

\end{document}